\documentclass{elsart}

\usepackage{epsfig}

\usepackage{amssymb}
\usepackage{amsmath}

\begin{document}

\begin{frontmatter}



\title{Mach cones and dijets - jet quenching and fireball expansion dynamics}


\author{Thorsten Renk}

\address{Department of Physics, PO Box 35 FIN-40014 University of Jyv\"askyl\"a, Finand}
\address{Helsinki Institut of Physics, PO Box 64 FIN-00014, University of Helsinki, Finland}

\begin{abstract}
The energy loss of a hard parton is regarded as a useful tool to probe properties of a soft medium. However, the information obtained by a measurement of the nuclear suppression $R_{AA}$ is not very sensitive to medium properties. Thus, more differential observables are needed. Here we focus on the pattern of angular correlation of hadrons associated with a hard trigger.
These show rich pattern when going from low $p_T$ to high $p_T$ associate momentum.  At low $p_T$, the pattern of associate hadrons can be understood in terms of Mach shocks excited by the energy lost from the away side parton while the parton itself is absorbed by the medium. At high $p_T$, fluid dynamics is not applicable any more and the correlations become dominated by the punchtrough of the away side parton with subsequent fragmentation. We present an analysis of these phenomena taking into account the full expansion dynamics of soft matter and show how both transverse and longitudinal expansion play a crucial role in understanding the data.
\end{abstract}

\end{frontmatter}

\section{Introduction}
\label{S-Intro}

The expression 'jet tomography' often used to describe the analysis of hard pQCD processes taking place inside the soft matter created in an ultrarelativistic heavy-ion collision. In particular the focus is on the nuclear suppression of hard hadrons in A-A collisions compared with the scaled expectation from p-p collisions due to loss of energy from the hard parton by interactions with the soft medium (see e.g. \cite{Tomo}). 
However, the nuclear suppression factor 
\begin{equation}
R_{AA}(p_T,y) = \frac{d^2N^{AA}/dp_Tdy}{T_{AA}({\bf b}) d^2 \sigma^{NN}/dp_Tdy}.
\end{equation}
is a rather integral quantity, arising in model calculations from the schematical convolution of the hard pQCD vacuum cross section $d\sigma_{vac}^{AA \rightarrow f +X}$ for the production of a  parton $f$, the energy loss probability $P_f(\Delta E)$ given the vertex position and path through the medium and the vacuum fragmentation function $D_{f \rightarrow h}^{vac}(z, \mu_F^2)$ as 
\begin{equation}
\label{E-Folding}
d\sigma_{med}^{AA\rightarrow h+X} = \sum_f d\sigma_{vac}^{AA \rightarrow f +X} \otimes P_f(\Delta E) \otimes
D_{f \rightarrow h}^{vac}(z, \mu_F^2),
\end{equation}
where
\begin{equation}
d\sigma_{vac}^{AA \rightarrow f +X} = \sum_{ijk} f_{i/A}(x_1,Q^2) \otimes f_{j/A}(x_2, Q^2) \otimes \hat{\sigma}_{ij 
\rightarrow f+k}.
\end{equation}
Here, $f_{i/A}(x, Q^2)$ denotes the nuclear parton distribution function dependent on the parton
momentum fraction $x$ and the hard momentum scale $Q^2$ and $\hat{\sigma}_{ij\rightarrow f+k}$ is the the partonic pQCD cross section.

Eq.~(\ref{E-Folding}) has to be properly averaged over all possible vertices distributed according to the nuclear overlap $T_{AA}$ and paths through the medium. In \cite{gamma-h} we have argued that it is possible to factorize this spatial averaging from the momentum space formulation Eq.~(\ref{E-Folding}) and thus define the geometry-averaged energy loss probability $\langle P(\Delta E, E) \rangle_{T_{AA}}$. $R_{AA}$ can thus be viewed as providing constraints for the form of  $\langle P(\Delta E, E) \rangle_{T_{AA}}$.

\begin{figure*}[htb]
\epsfig{file=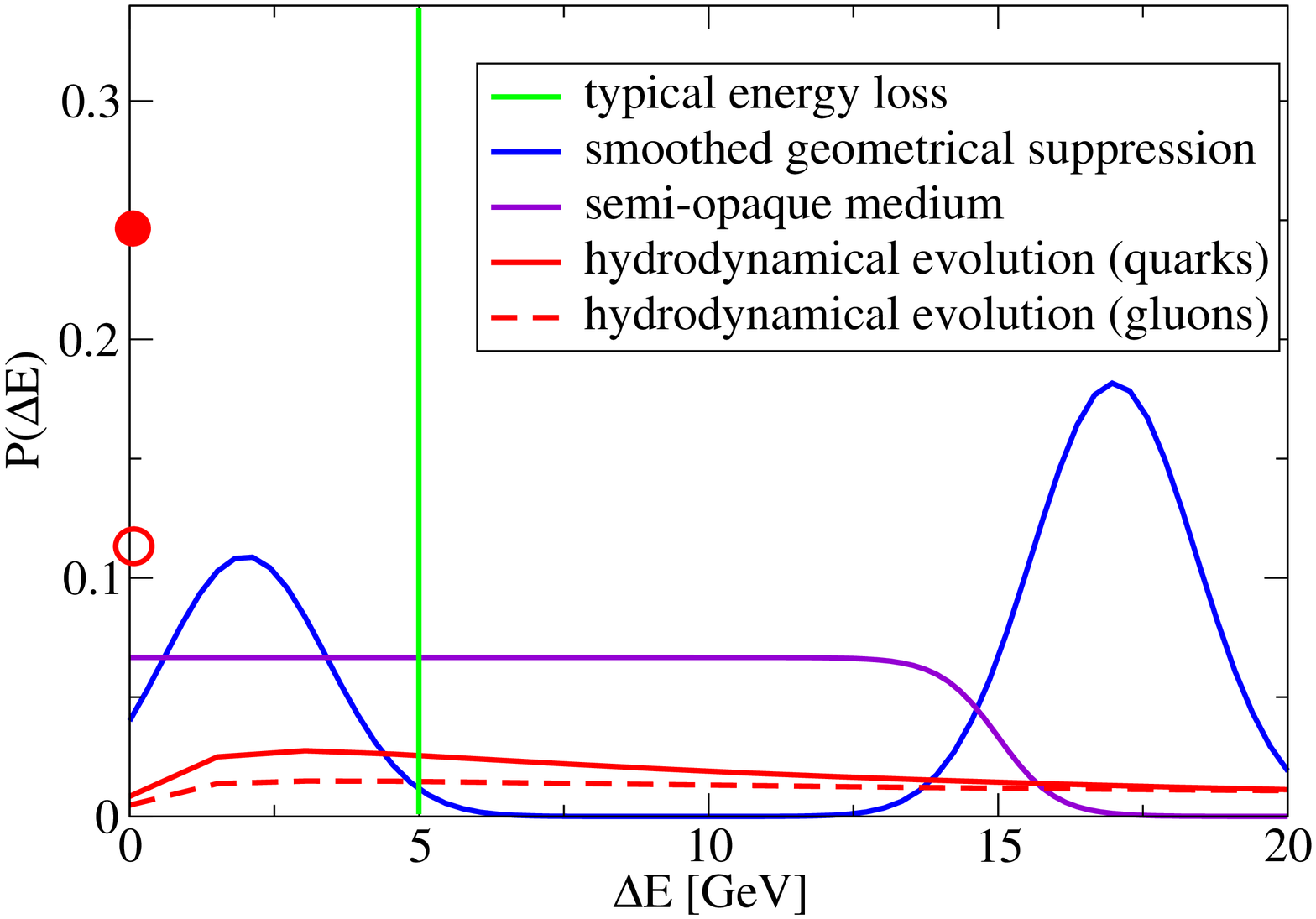, width=6.9cm} \epsfig{file=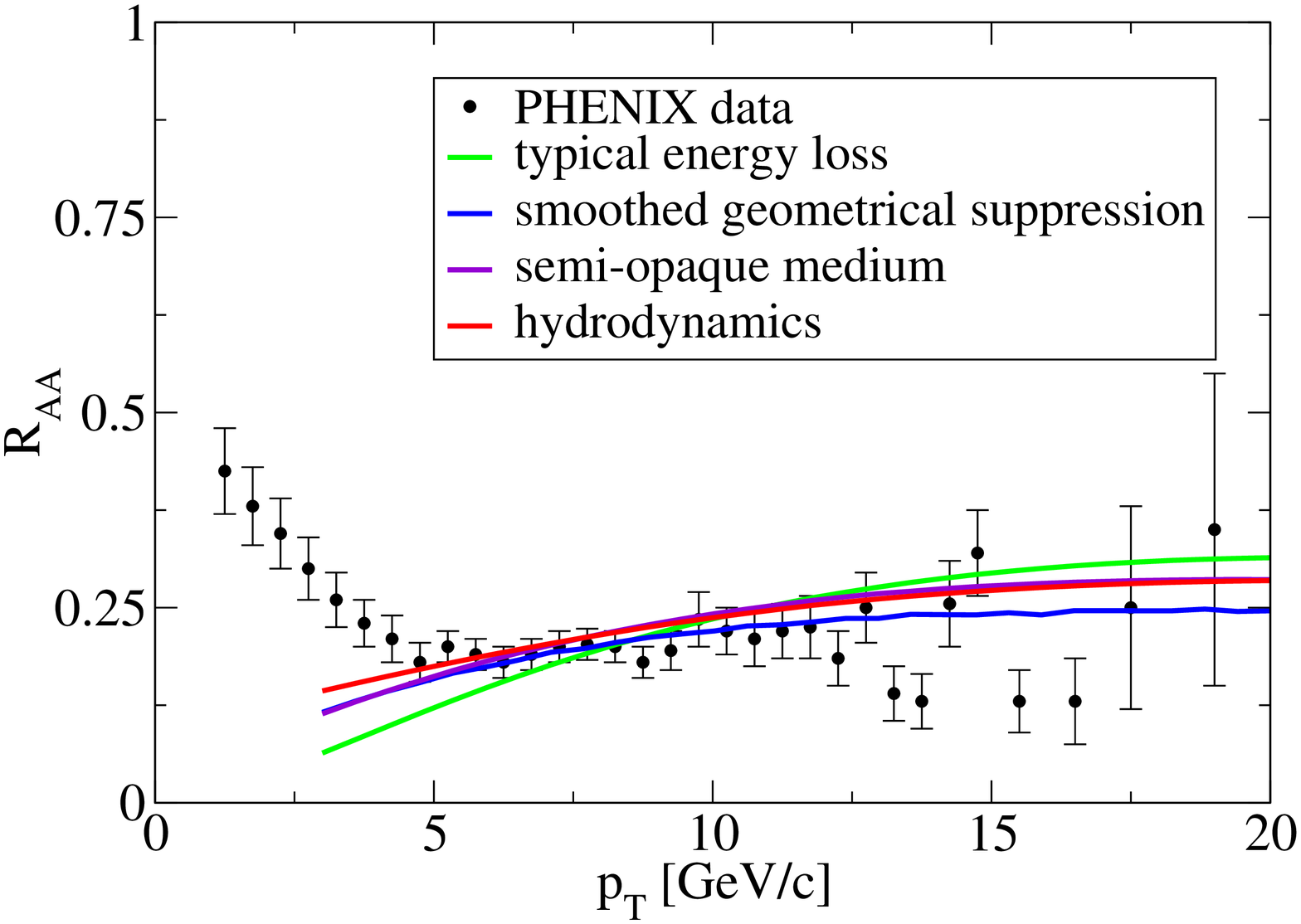, width=6.9cm}
\caption{\label{F-Comp}Left panel: Geometry-averaged energy loss probability distributions for different energy loss scenarios\cite{gamma-h}. Right: Nuclear suppression factor $R_{AA}$ for the different scenarios compared with PHENIX data \cite{PHENIX_R_AA}.}
\end{figure*}

In Fig.~\ref{F-Comp} we show trial distributions $\langle P(\Delta E, E) \rangle_{T_{AA}}$ based on different possible scenarios how the medium induces jet quenching. Although the form of the individual $\langle P(\Delta E, E) \rangle_{T_{AA}}$ differs substantially, the resulting $R_{AA}$ is very similar. This implies that $R_{AA}$ does not provide strong enough constraints to extract medium properties reliably except for the fact that quenching is substantial \cite{gamma-h}. However, given that assumptions about the longitudinal and transverse flow profile (while requiring that the soft hadron spectra agree with data) can alter the medium quenching power by a factor of five \cite{JetFlow}, this means that any attempt to gain access to information about the medium density must involve observables beyond $R_{AA}$. In \cite{gamma-h} we demonstrated that $\gamma$-hadron correlations \cite{XNPhotons} are capable of resolving the differences; in this paper we're interested in the additional information accessible by angular correlation measurements.

\section{The framework}

Energy and momentum lost by a hard parton reappear in the medium. We make the assumption that the bulk of the lost energy excites a shockwave in the soft medium. If we study back-to-back production of hard partons and require one of the partons to yield a hard hadron trigger, the angular correlation pattern associated with this trigger consists of two components with distinct properties. First, there are the remnants of the hard processes --- next to leading hadrons created in the fragmentation alongside the trigger and hadrons created in the fragmentation of the away side parton whenever this traverses the medium. The only place medium properties appear in this contribution is the amount of energy lost from the partons, but as long as fragmentation occurs in vacuum, neither correlation angular position nor width is determined by the medium, only the correlation strength.
In contrast, of the soft process (i.e. the excited shockwave) only the strength is not determined by the medium but by the lost energy. The position is largely determined by the medium speed of sound $c_s$ \cite{Shuryak,Mach} and its width by the freeze-out conditions \cite{Mach}.

We describe the hard underlying process by Monte-Carlo sampling Eq.~(\ref{E-Folding}) for the near and away side. The energy loss probability $P(\Delta E)$ given the path is determined by using BDMPS radiative energy loss in the formulation of \cite{QuenchingWeights} by evaluating line integrals over the local quenching power along the path of the hard parton \cite{JetFlow,Dijets}. As a description of the medium we mainly consider a parametrized evolution capable of describing a plethora of bulk observables \cite{Parametrized} but we also study a hydrodynamical evolution \cite{Hydro}. 

In order to simulate the shockwave we follow the flow of energy and momentum. We assume that a fraction $f$ of the energy and momentum lost to the medium 
excites a shockwave characterized by a dispersion relation $E = c_s^2 p$ and a fraction
$(1-f)$ in essence heats the medium and leads to collective drift along the jet
axis to conserve momentum \cite{Shuryak}.
We calculate $c_s$ from a quasiparticle description of the equation of state.
The dispersion relation along with the energy and momentum deposition determines the initial angle of propagation
of the shock front with the jet axis (the 'Mach angle') as $\phi = \arccos c_s$. We discretize the time into small intervals $\Delta \tau$, calculate the energy deposited in that time as
$E(\tau) = \Delta \tau \cdot dE/d\tau$ and then propagate the part of the shockfront remaining in the midrapidity
slice (i.e. in the detector acceptance). Each piece of the front is propagated with the local speed of sound and the angle of propagation is
constantly adjusted as 
$
\phi = \arccos {\int_{\tau_E}^\tau c_s(\tau) d\tau }/{(\tau - \tau_E)}.
$
Once an element of the wavefront reaches the freeze-out condition $T = T_F$, a hydrodynamical
mode cannot propagate further.  In the local
restframe, we have a matching condition for the dispersion relations
$E = c_s^2 p$ and $E = \sqrt{M^2 + p^2} - M$
where $M = V \left(p(T_F) + \epsilon(T_F)\right)$ is the 'mass' of a volume element at $T_F$.
Once we have calculated the additional boost $u_\mu^{shock}$ a volume element receives from the shockwave using the
matching conditions, we use the Cooper-Frye formula
to convert the fluid element into a distribution of hadrons.

\section{Low $p_T$ associate --- recoil of the hydro medium}

\begin{figure}[htb]
\epsfig{file=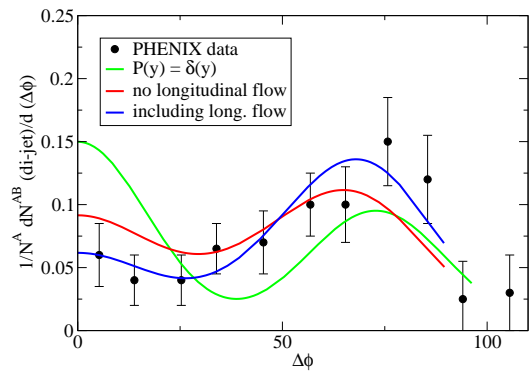, width=6.9cm}
\caption{\label{F-Elongation}Calculated 2-particle correlation under the assumption that a) the away side parton is always at midrapidity and the excited mode doesn't couple to flow (green) b) using realistic $P(y)$ and assuming that the excited mode doesn't couple to flow (red) and c) including realistic $P(y)$ and longitudinal flow elongation compared with PHENIX data \cite{PHENIX-2pc}. }
\end{figure}

At an associate cut of about 1 GeV, the system is clearly dominated by the recoil of the hydro medium, we observe almost no away side partons emerging from the medium in the simulation. The shockwave shows a strong influence on longitudinal and transverse flow: Due to the fact that the rapidity $y$ of the away side parton is not fixed by the rapidity of the trigger hadron, an averaging over $y$ leads to an apparent shrinking of the cone \cite{MachRap}, cf. Fig.~\ref{F-Elongation}. For a scenario which does not couple to longitudinal flow, this effect leads to disagreement with the data.

However, for a hydro mode the shock cone is longitudinally elongated due to its coupling to flow, therefore its spatial extension is the solution of the characteristic equation
\begin{eqnarray}
\left.\frac{dz}{dt}\right|_{z=z(t)}=\left.\frac{u(z,R,t)+c_s(T(z,R,T))}{1+u(z,R,t)c_s(T(z,R,t))}\right|_{z=z(t)}
\end{eqnarray}
and the subsequent widening in rapidity space leads to agreement with the data. Any scenario which does not couple to longitudinal flow must thus necessarily exhibit an even wider angle before averaging. 

\begin{figure*}[!htb]
\epsfig{file=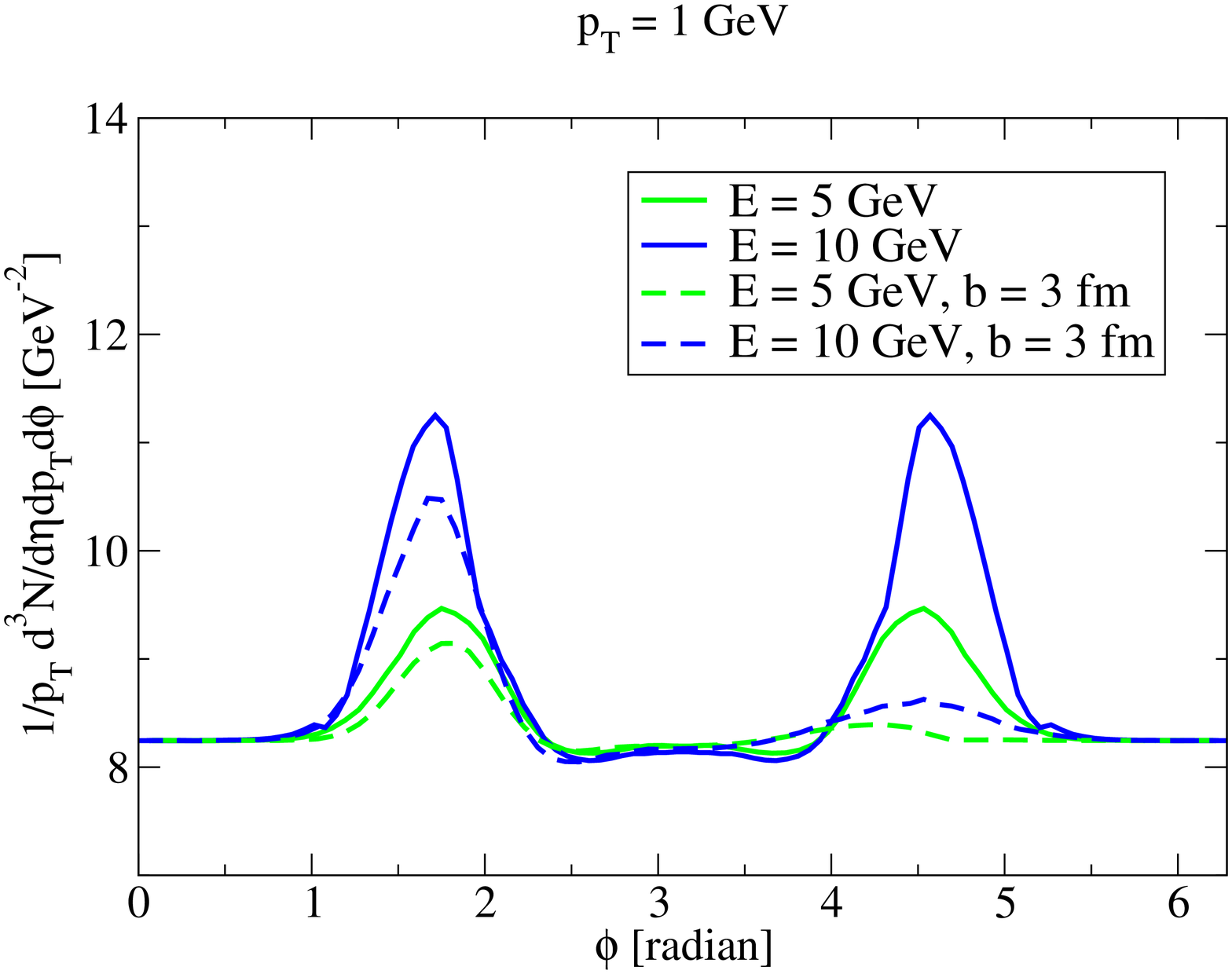, width=6.9cm}\epsfig{file=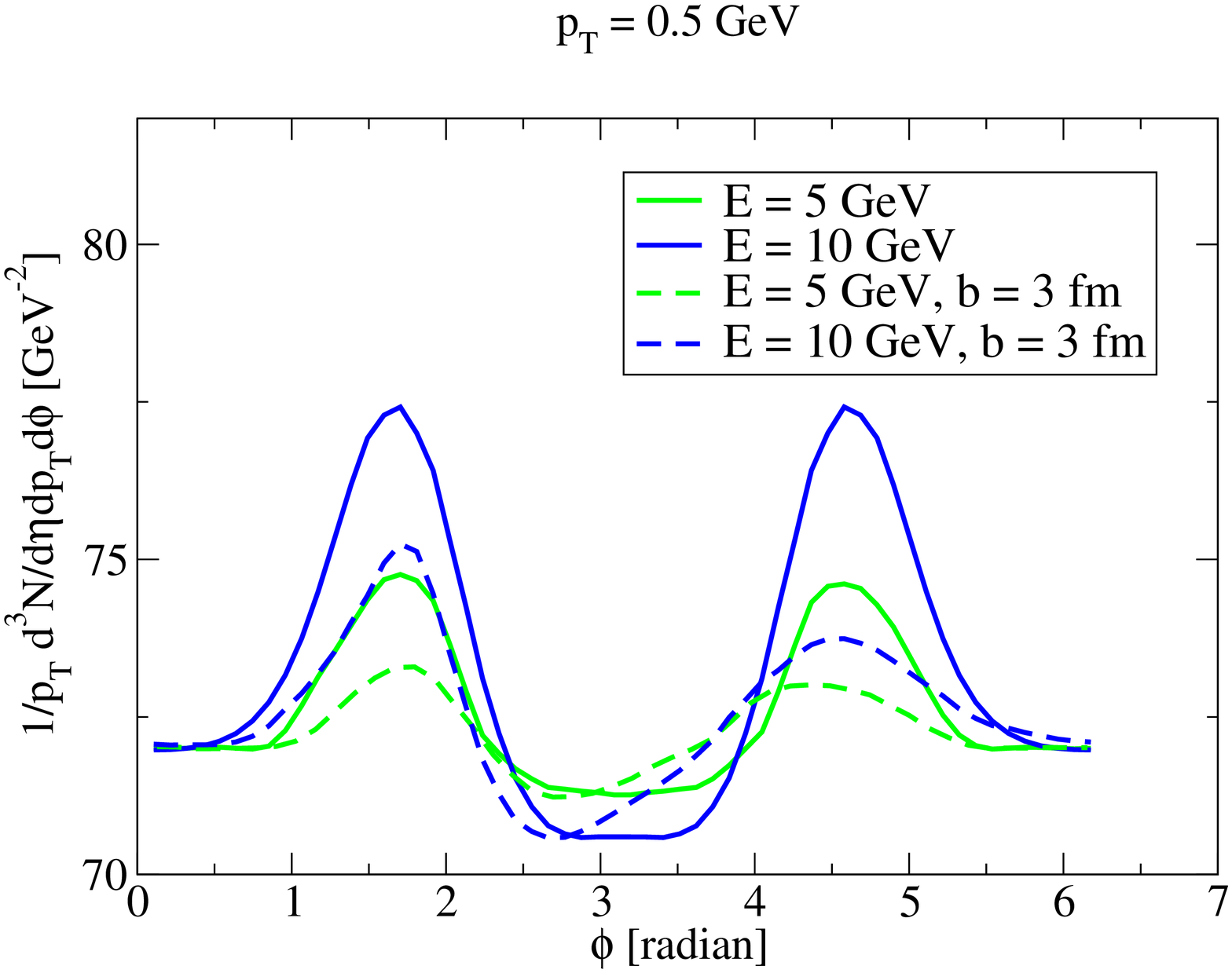, width=6.9cm}
\caption{\label{F-Disappearance}The effect of transverse flow on the observable correlation peak for configurations in which transverse flow 
and cone
are aligned (solid) and orthogonal (dashed, $b=3$ fm) for associate hadron $p_T = 1.0$ GeV 
(left panel) and $500$ MeV (right panel). All calculations are for fixed energy $E$, intrinsic $k_T=0$ and $y_{away}=0$. Correlation strength in the 
direction of the away side parton is suppressed for clarity. }
\end{figure*}

Transverse flow has no great influence on the position of the angle, but due to the fact that high $p_T$ correlations in a hydro + shockwave description are only generated when flow and shockwave are almost aligned, it may lead to the apparent disappearance of one wing of the shockwave at sufficiently high $p_T$ (which is recovered at lower $p_T$ for some initial vertices. We illustrate this in Fig.~\ref{F-Disappearance}. This has considerable impact on the interpretation of 3-particle correlation data.

\section{Raising $p_T$ --- competition between hard and soft mechanisms}

It is clear that the relative importance of the hard and soft component of the correlation will depend on both trigger threshold and associate hadron cut. As soon as hydrodynamic flow of thermalized matter ceases to be the chief mechanism of hadron preduction at given $p_T$, it cannot longer be assumed that shockwaves dominate the correlations. Even for a 20 GeV trigger hadron, one cannot expect to find shockwave correlations at 6 GeV, simply because there is no hydro medium to support them at this scale. Thus, the rising associate cut will in general shift the focus from soft to hard modes.
The expected effect of a rising trigger is somewhat more complex. Increasing the parton energy will increase the likelyhood of observing punchthrough, but also the yield on the near side (caused by next to leading fragmentation) as compared to the away side (cause by shockwaves) will change with the excitation function of these mechanisms with energy. Hard processes are expected to dominate at large enough scales.

Both the vanishing of the low $p_T$ dip (possibly due to punchthrough) and the drop of the away side orrelation strength compared to the near side expected from the model are qualitatively seen in the data \cite{STAR1} for increased trigger and associate momentum scale.
Especially the fact that the away side pattern remains broad for increasing associate momentum and only drops in strength is a strong point in favour of a hydrodynamical mode, as pQCD motivated emission scenarios predict a shrinking of the angle.
 However quantitative regions in the transition regime between hard and soft physics are challenging and have not seriously been attempted so far. 

\section{Hard trigger and hard associate --- punchthrough} 

\begin{figure*}[htb]
\epsfig{file=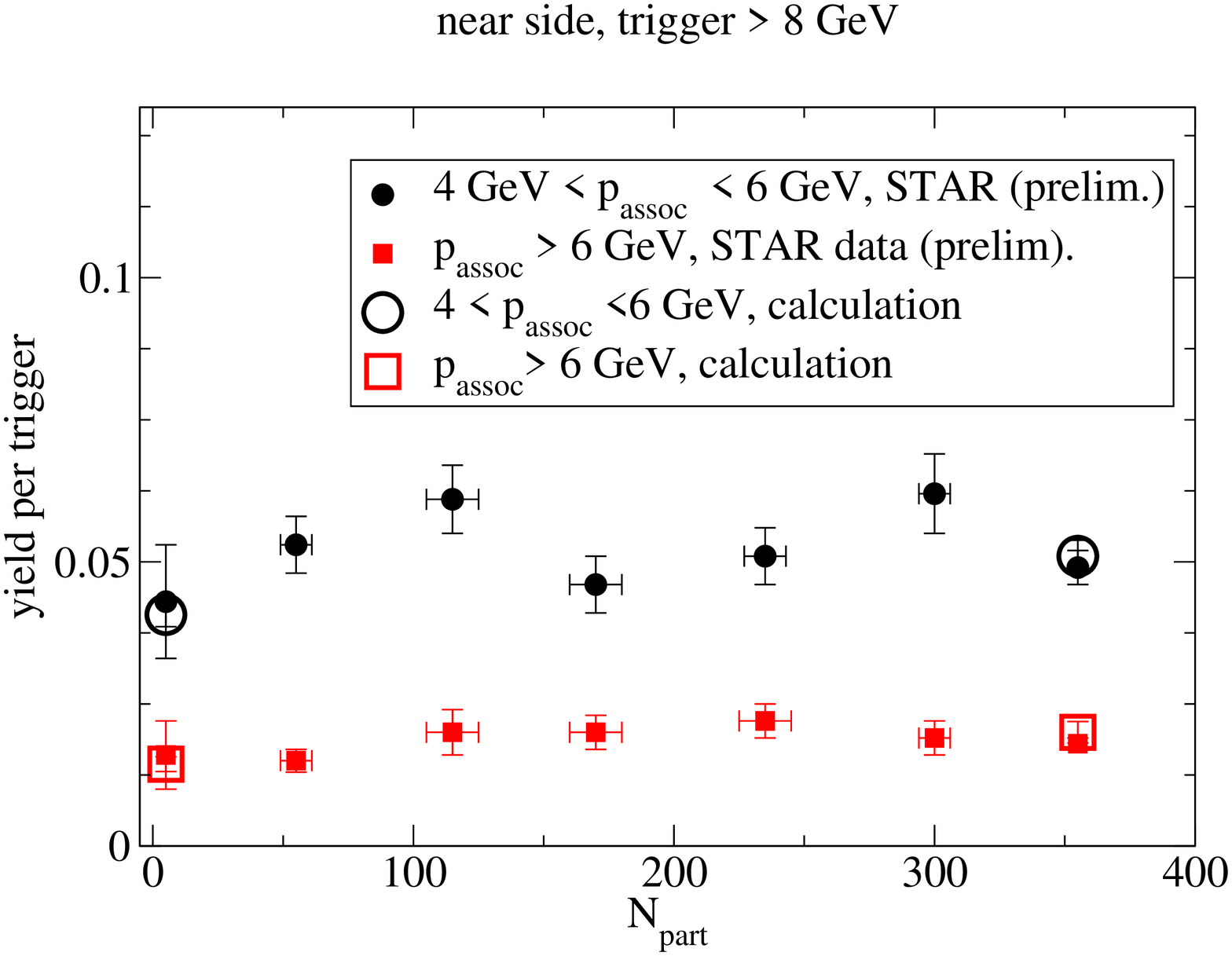, width=6.8cm}\epsfig{file=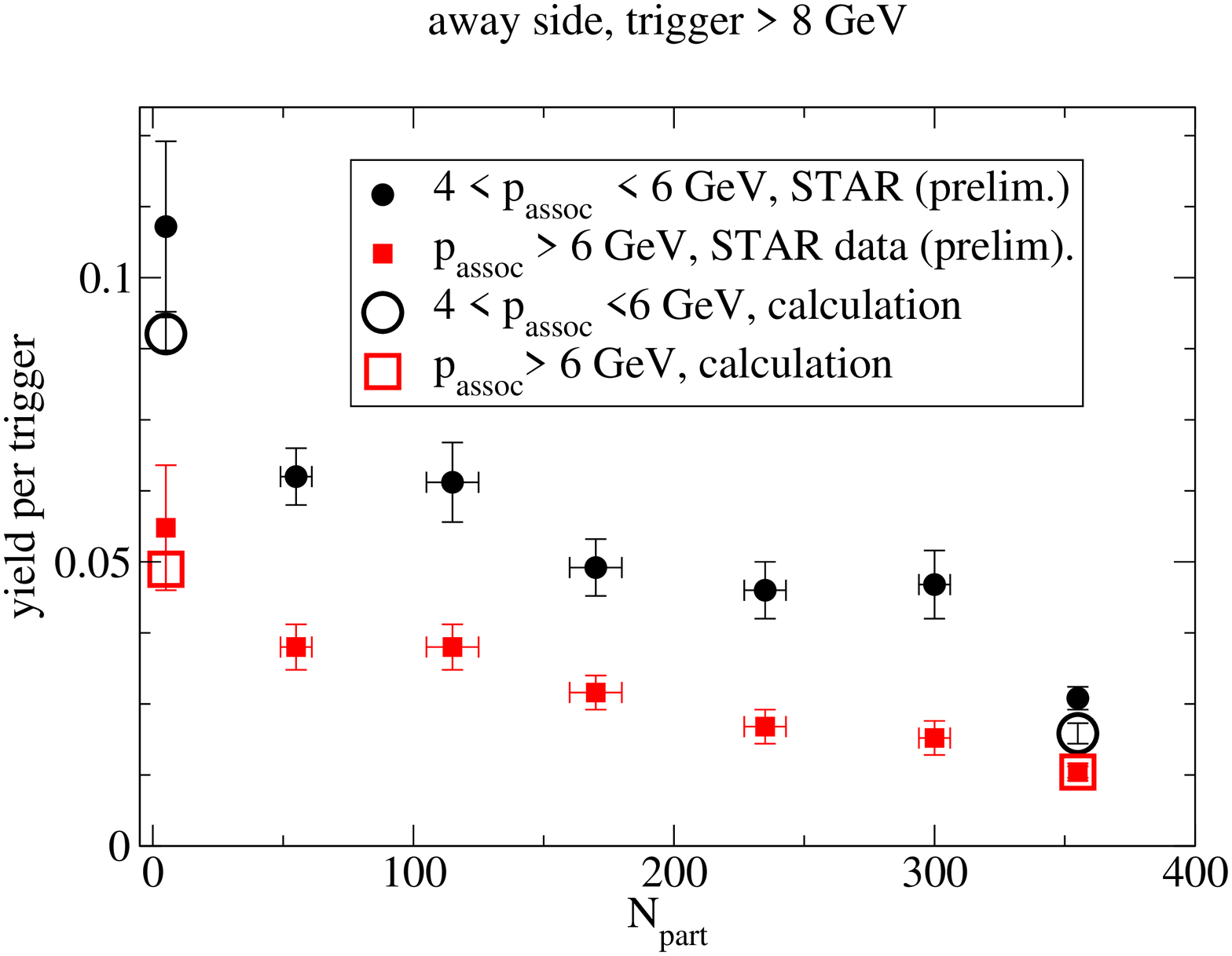, width=6.8cm}
\caption{\label{F-yield}Yield per trigger on the near side (left panel) and away side (right panel) for trigger hadron above 8 GeV in the model calculation as compared with STAR data \cite{DijetsSTAR}.}
\end{figure*}

If the trigger $p_T$ is raised in the region above 6 GeV, punchthrough of the away side parton through the expanding medium is observed, and the yields obtained in the model agree well with the data, cf. Fig.~\ref{F-yield}, except in the 4-6 GeV away side momentum bin. This indicates that the hadron production in this region in the presence of a medium is not primarily given by fragmentation of a hard parton but that sizeable contributions of other processes are present, quite possibly recombination \cite{Reco} plays a role here.

\begin{figure*}[htb]
\epsfig{file=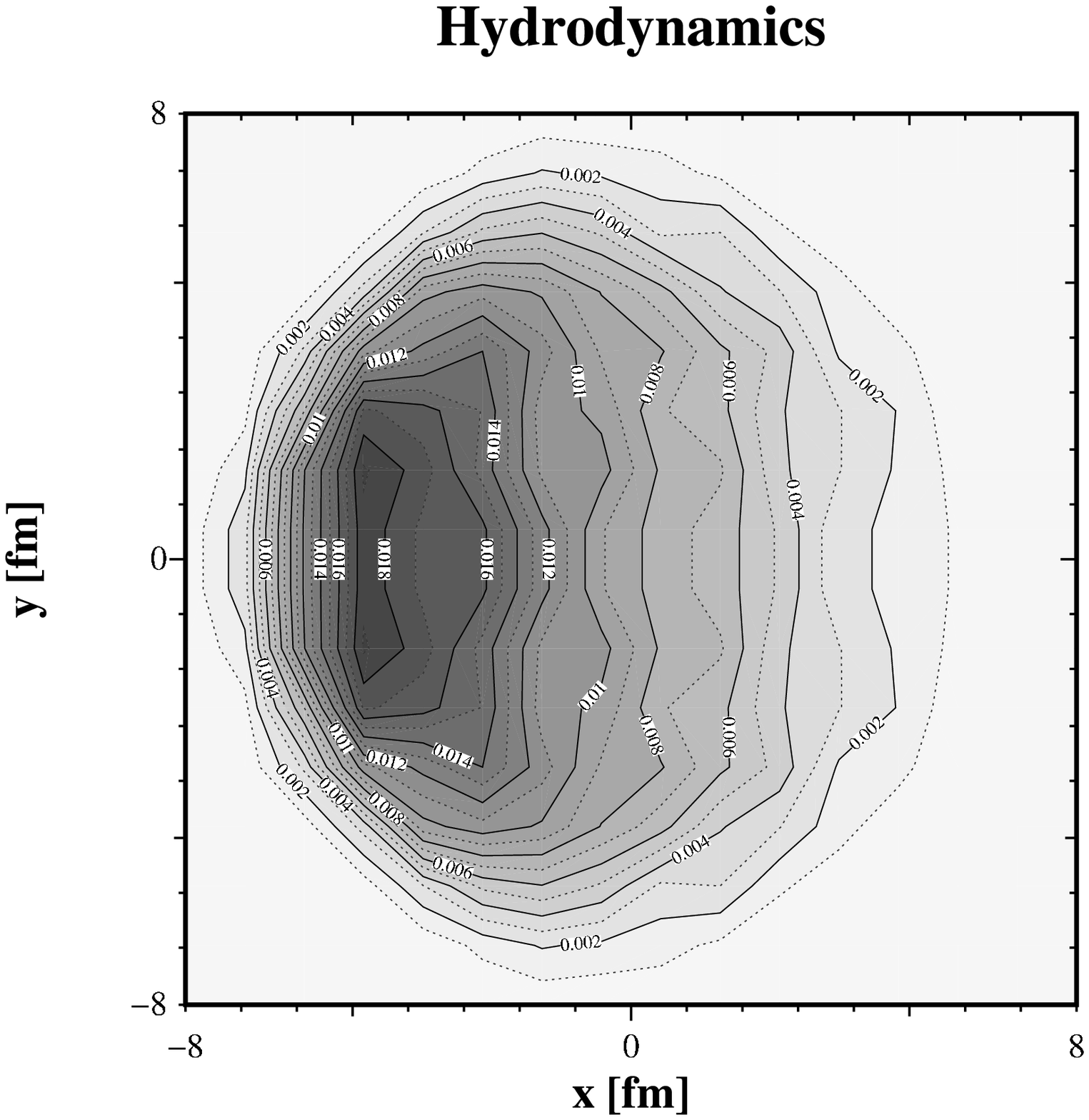, width=6.4cm}\epsfig{file=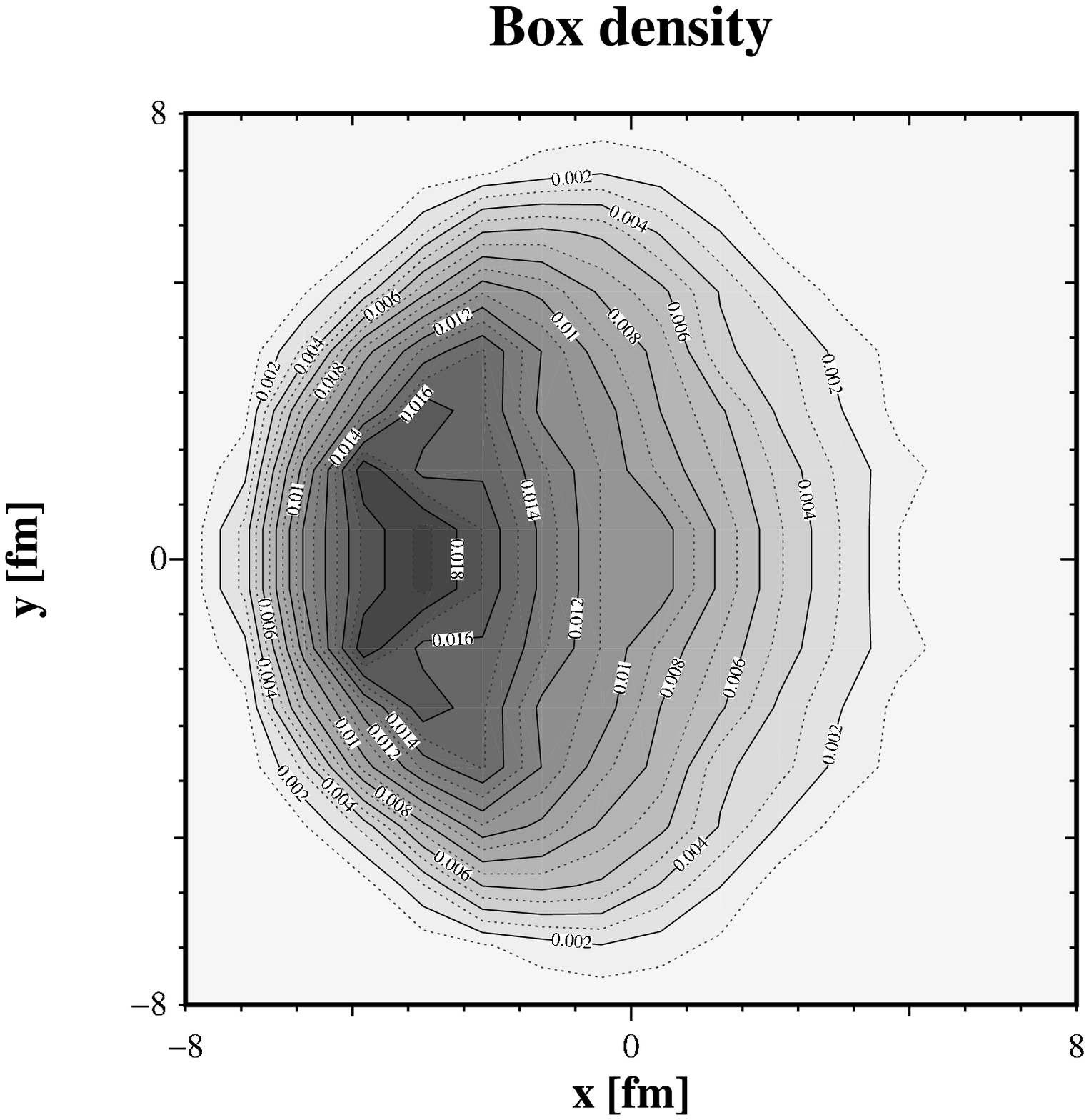, width=6.4cm}\\
\vspace*{-2.7cm}
\caption{\label{F-vertex}Normalized density of vertices for events with a trigger hadron above 8 GeV. The near side parton propagates into negative x direction.}
\end{figure*}

In Fig.~\ref{F-vertex} we show the distribution of vertices leading to a triggered hadron (note that this distribution is also relevant for $R_{AA}$. While there is some degree of surface bias visible, the core of the system is by no means black and contributions to $R_{AA}$ can come also from the core of the matter. This is in contrast to previous studies \cite{Fragility} and the difference cen be traced back to the fact that the full expansion is taken into account in the present model \cite{Correlations}.

\begin{figure*}[htb]
\epsfig{file=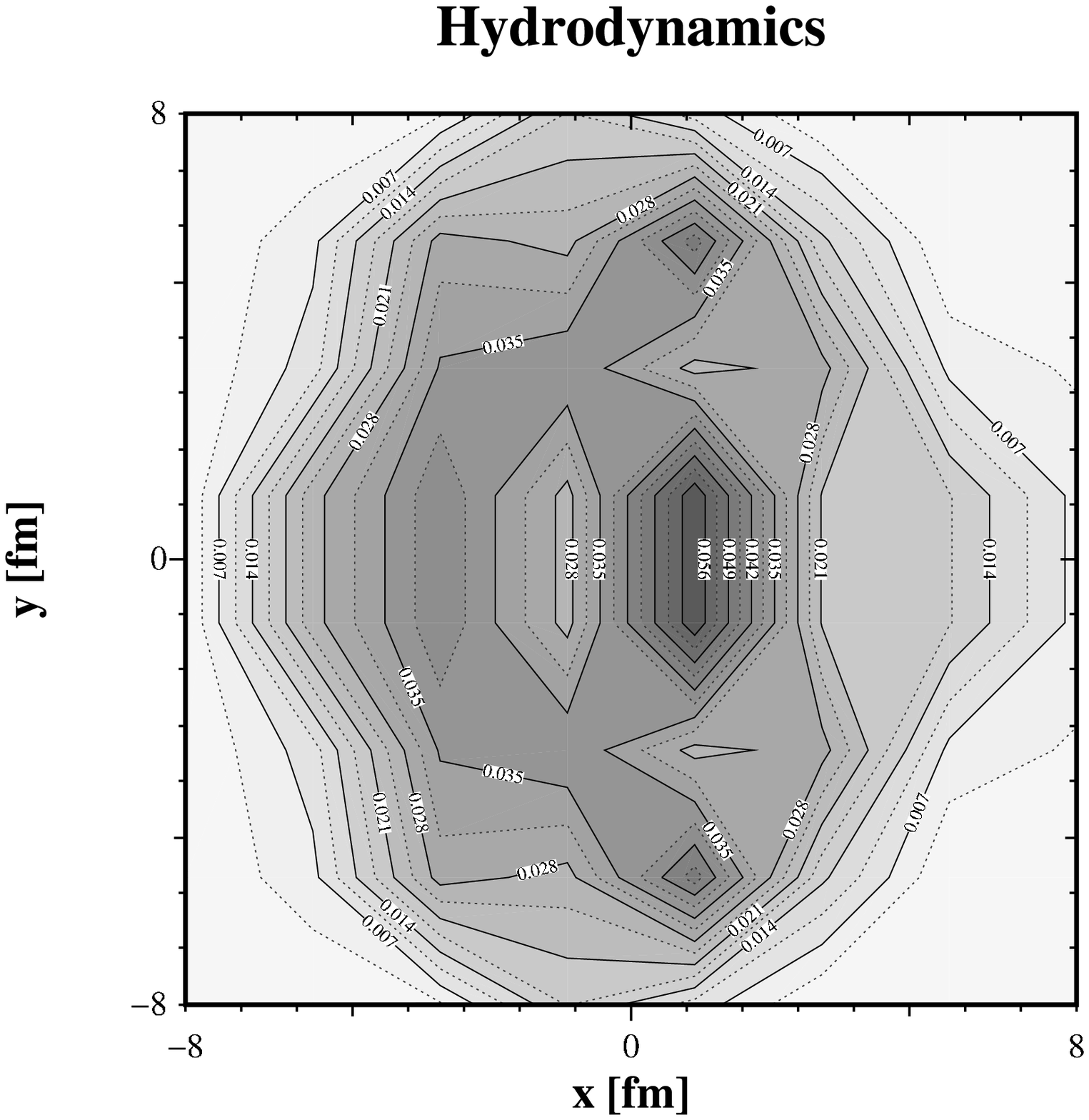, width=6.4cm}\epsfig{file=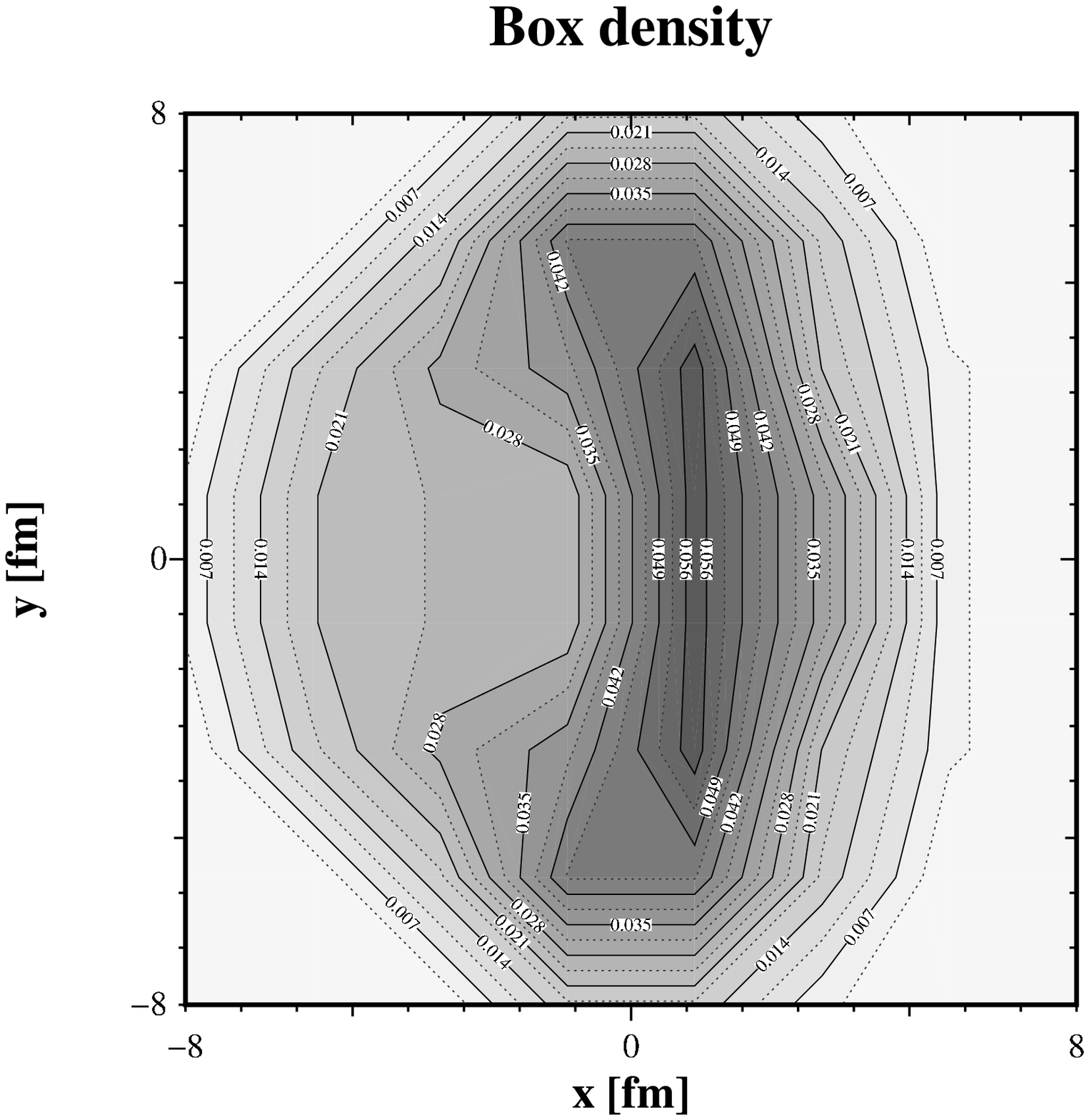, width=6.4cm}\\
\vspace*{-2.7cm}
\caption{\label{F-pvertex} Normalized density of vertices for which an associate hadron in the momentum region 4 $< p_T <$  GeV is found, given a trigger in the region 8 $< p_T <$ 15 GeV. The near side parton propagates into negative x direction.}
\end{figure*}

In Fig.~\ref{F-pvertex} we illustrate the spatial region probed in back-to-back dihadron measurements. This region shows rather large model-dependence, but is in general peaked more towards the medium center. This geometry dependence is a promising handle to gain information about the medium density distribution from dihadron correlation measurements \cite{Correlations}.

\section{Summary}

We have argued that $R_{AA}$ is not sufficiently sensitive to medium properties and that more differential measurements are needed. We have presented a consistent picture of angular correlation pattern given a hard hadron trigger based on hard processes and the recoil of the soft medium. Within this unified description, the correlations probe both the medium flow profile and the density distribution, thus in principle more detailed information can be gained. In practice, more high-precision data are necessary --- while it is easy to rule out toy models, the sensitivity to distinguish different dedicated bulk matter models is not yet sufficient \cite{Correlations}.



\end{document}